\documentclass[%
 reprint,
superscriptaddress,
 amsmath,amssymb,
 aps,
 pra,
 floatfix
]{revtex4-1}

\usepackage{color,soul}
\usepackage{float}
\usepackage{graphicx}% Include figure files
\usepackage[dvipsnames]{xcolor}
\usepackage{dcolumn}% Align table columns on decimal point
\usepackage{bm}% bold math
\usepackage{braket}%for bra-ket notation
\usepackage{bbold}
\usepackage[unicode]{hyperref}
\usepackage[utf8]{inputenc}%ä,ö,ü
\usepackage{cleveref}
\crefname{equation}{}{}
\Crefname{equation}{}{}
\usepackage{subcaption}
\usepackage{siunitx}
\usepackage{float}
\newcommand*{\new}{\textcolor{black}}
\usepackage{array}
\usepackage{microtype}
\usepackage{appendix}

\usepackage{amsmath}
\usepackage{mathtools}

\begin{document}

\preprint{APS/123-QED}

\title{Generation and Robustness of Quantum Entanglement in Spin Graphs}

\author{Jan Riegelmeyer}
\affiliation{Department of Physics, University of York, York, YO10 5DD, United Kingdom}
\affiliation{Fachbereich Physik, Westfälische Wilhelms-Universität Münster, 48149 Münster, Germany}
\author{Dan Wignall}
\affiliation{Department of Physics, University of York, York, YO10 5DD, United Kingdom}
\author{Marta P. Estarellas}
\affiliation{National Institute of Informatics, 2-1-2 Hitotsubashi, Chiyoda-ku, Tokyo 101-8430, Japan}
\author{Irene D'Amico}
\affiliation{Department of Physics, University of York, York, YO10 5DD, United Kingdom}
\author{Timothy P. Spiller}
\affiliation{Department of Physics, University of York, York, YO10 5DD, United Kingdom}

%\date{\today}

\begin{abstract}

Entanglement is a crucial resource for quantum information processing, and so protocols to generate high fidelity entangled states on various hardware platforms are in demand. While spin chains have been extensively studied to generate entanglement, graph structures also have such potential; however, only a few classes of graphs have been explored for this specific task. In this paper, we apply a particular coupling scheme involving two different coupling strengths to a graph of two interconnected $3\times3$ square graphs such that it effectively contains three defects. We show how this structure allows generation of a Bell state whose fidelity depends on the chosen coupling ratio. We apply partitioned graph theory in order to reduce the dimension of the graph and show that, using a reduced graph or a reduced chain, we can still simulate the same protocol with identical dynamics. Finally, we investigate how fabrication errors affect the entanglement generation protocol and how the different equivalent structures are affected, finding that for some specific coupling ratios they are extremely robust.

\end{abstract}
\maketitle

\section{Introduction}

\new{Quantum computers hold the promise of being one of the next major technological developments in the field of information technology \cite{nielsen,deutsch1992rapid,ladd2010quantum}. Quantum phenomena, such as entanglement and superposition of states, provide quantum computers with the ability to potentially solve some hard computational problems and to simulate quantum systems in a more efficient way than their classical counterparts \cite{shor1997,arute2019quantum,bennett2000quantum,bruss2007lectures}. However, one of the current limitations of this technology relies on the number of qubits that can be allocated in a single chip} \cite{bose2007,Kielpinski2002}. The impact of this issue can be alleviated by connecting different chips or registers through a quantum bus \cite{bose2007,Skinner2003}. When these interconnections are relatively short, it is desirable to use the same physical platform and avoid using hybrid systems due to the associated inter-conversion from and to different encoding degrees of freedom (e.g. states of light in optical links) \cite{Kielpinski2002,bose2007}. For that purpose, arrangements (chains or graphs) of solid-state qubits are good candidates for short-range communication \cite{bose2007,Skinner2003,Kielpinski2002,bohnet2016quantum}. In addition to their application as quantum buses, spin chains and graphs are also able to perform other quantum information processing tasks, such as the creation and distribution of an entangled state \cite{spiller2007,sahling2015,Estarellas2017}. 

\new{The use of direct physical links is not the only way of transferring quantum states and, for example, the teleportation protocol proposed by Bennett \cite{bennett1993} uses entanglement to communicate quantum information.} Entanglement is also present in a wide range of other applications, such as one-way quantum computer architectures \cite{raussendorf2001,walther2005} or quantum key distribution \cite{ekert1991quantum,Jennewein2000,Bennett1992}. Given that entanglement is a useful resource for many applications, a reliable way to generate distributed entangled states on demand is paramount.

\new{In this paper, we explore the dynamics and entanglement generation/distribution capabilities of a spin graph formed by two interconnected $3\times3$ square graphs. This is engineered to present an `ABC-coupling' configuration \cite{Estarellas2017} with different coupling ratios.} We use the methods of graph partitioning from Refs.~\cite{kay2018,Bachmann2012} to simplify such graph into a quotient graph and a quotient linear chain. In Sect.~\ref{sec:mod}, we explain in detail the spin \new{graph} model. \new{We also present the structure of the spin graph under study and explain the partitioning theory that allows its simplification. In addition, we introduce the measure used to assess the quality of entanglement (the entanglement of formation, or EOF).} In Sect.~\ref{sec:res} we present our results. For different coupling ratios, we compare the values of EOF obtained in a short period of time (\new{something relevant for experiments on hardware with short decoherence times}) against the maximum EOF values over a larger time window. We then investigate the effects that fabrication errors have on the entanglement generation. We analyse both the effects of errors on the couplings between qubits (non-diagonal disorder) and of errors on the on-site qubits' energies (diagonal disorder). Finally, our conclusions are included in Sect.~\ref{sec:concl}.

\section{\label{sec:mod}The Model}
\new{Let us now explain the main concepts behind our spin graphs and the techniques that are here being used to analyse their entanglement generation properties. In the following subsections we provide a general introduction to spin graphs and explain the details on the partitioned graph theory. We then focus on the graph structure under study and introduce the `ABC-coupling' configuration. We show that with this structure and coupling scheme, and the right tuning of the coupling parameters, the natural dynamics of the system can generate a distributed highly entangled state.}

%structure/design: Partitioned graph theory
%couplings: ABC-coupling configuration

\subsection{\label{sec:spinchain}The spin \new{graph} formalism}

\begin{figure}[h!]
\centering
\resizebox{0.45\textwidth}{!}{
  \includegraphics{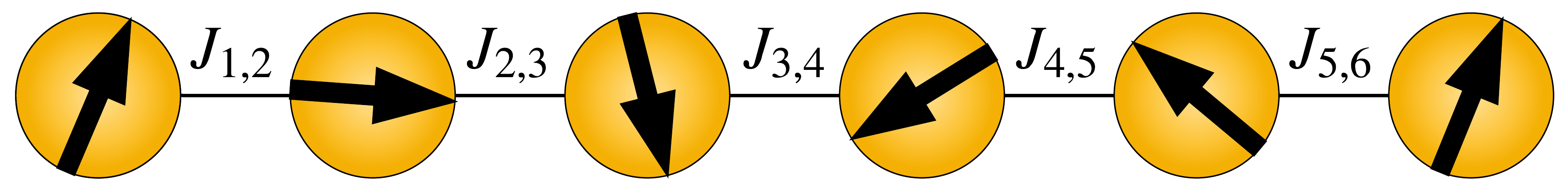}}
\caption{Diagram of a spin chain of 6 qubits. $J_{i,i+1}$ are the coupling energies between two adjacent qubits. Tilted arrows pictorially represent superpositions of up and down spin states.}
\label{fig1}       
\end{figure}

We describe generic spin graph systems with the XY (sometimes also called XX) Heisenberg Hamiltonian. With $\ket{0}$ and $\ket{1}$ as our $\sigma_{z}$ basis states, we write the Hamiltonian as
\begin{eqnarray}
H_{XY}&= &\sum_{i=1}^{N-1}\sum_{\mathclap{\substack{j>i}}}^{N}J_{i,j}[\ket{1}\braket{0|_i\otimes |0}\braket{1|_{j}+|0}\braket{1|_i\otimes|1}\bra{0}_{j}]\nonumber\\
& &+\sum_{i=1}^N \epsilon_i\ket{1}\bra{1}_i,
    \label{eq2}
\end{eqnarray}
where $N$ is the number of qubits of the graph and $J_{i,j}$  is the coupling energy between qubit $i$ and qubit $j$. We will consider all the energies $\epsilon_i$ to be equal, time-independent and scaled to zero unless otherwise stated. An example of a simple spin graph is the one-dimensional spin chain illustrated in Fig.~\ref{fig1}.

 As already noted, some particular arrangements of spin graphs present good quantum state transfer properties \cite{estarellas2018,bose2007,christandl2004,kay2010perfect}, meaning that a quantum state can be reliably transmitted from a specific qubit of the graph (sender) to another (receiver). 
 
 \subsection{\label{sec:PartGraph}Partitioned Graph Theory}

 Here, we consider the theory of graph partitioning based on \cite{kay2018}. This allows reducing the complexity of a graph by collapsing several similar sites into a single effective site that will be referred to as a node. The obtained graph is called the partitioned graph. After also readjusting the new interaction strengths one then obtains the quotient graph. The purpose of our use of partitioned graph theory is to find and study different graph structures with equivalent dynamics. This is not merely to provide a simpler calculation, but rather to present different graphs with the same entanglement generation properties. \new{This would allow experimentalists to be able to choose between different graph topologies and find the most suitable to the characteristics of the available hardware.} We will refer to the original graph, for which the complexity will be reduced by partitioning, as the full graph and call its sites \textit{fg-sites}. \new{All coupling energies in this graph are equal.} In \cite{kay2018}, a partitioned graph G comprising a set of nodes is defined such that:
 \begin{itemize}
       \item The first node comprises a single \textit{fg-site} of the full graph.
       \item All \textit{fg-sites} collapsed in node $i$ are equidistant from the first node.
       \item For any pair of nodes, $i, j$, every \textit{fg-site} collapsed in node $i$ connects to the same number of \textit{fg-sites} collapsed in node $j$. 
       \item No edges join \textit{fg-sites} collapsed in the same node.
    \end{itemize}

\new{Once the first node is defined, we will group the remaining \textit{fg-sites} together to form the other nodes of the partitioned graph. These remaining \textit{fg-sites} are grouped \emph{if and only if} they all have the same distance to the first node and the same coupling degree to the \textit{fg-sites} collapsed into a different node.} Note that the distance between two sites is defined by the number of edges connecting them on the shortest path.

\begin{figure}[h!]
        \centering
        \includegraphics[width=0.25\textwidth]{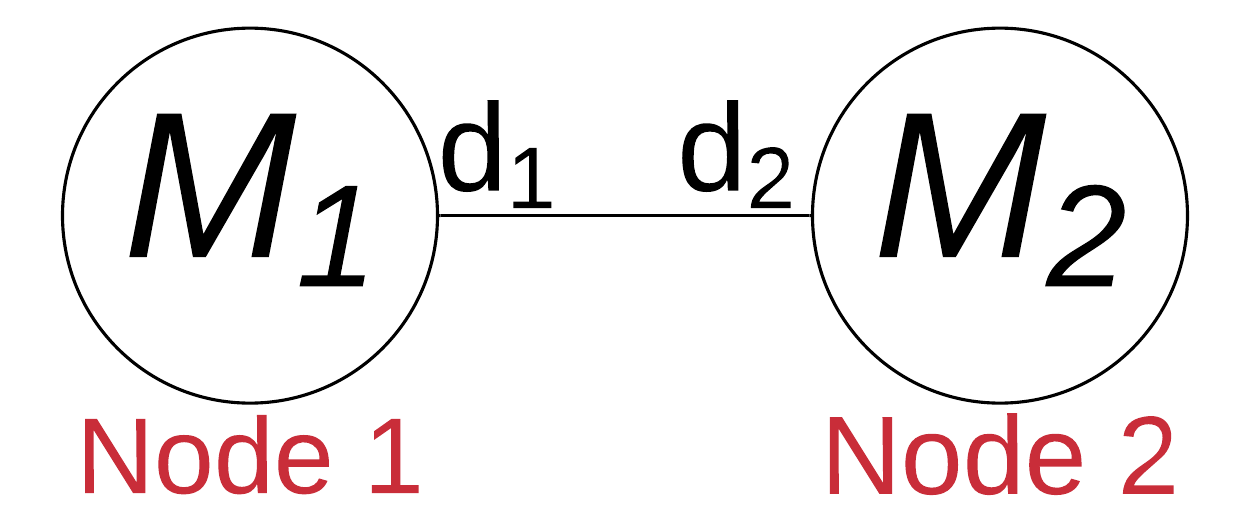}
        \caption{Example of a partitioned graph consisting of two nodes. Node 1 is connected to $d_1$ \textit{fg-sites} of node 2, and similarly node 2 is connected to $d_2$ \textit{fg-sites} of node 1. Node 1 comprises $M_1$ \textit{fg-sites} and node 2 comprises $M_2$ \textit{fg-sites} \cite{kay2018}.}
        \label{2}
\end{figure}

In Fig.~\ref{2}, we show two nodes, where each \textit{fg-site} of the full graph in node 1 is coupled to $d_1$ \textit{fg-sites} in node 2 and each \textit{fg-site} in node 2 is coupled to $d_2$ \textit{fg-sites} in node 1. \new{Here, $M_1$ ($M_2$) is the number of \textit{fg-sites} collapsed into node 1 (2)}. It is always required that $M_1d_1=M_2d_2$, or more generally 
\begin{equation}
   M_id_i=M_jd_j  
   \label{eq3}
\end{equation} for any coupled pair of nodes $i$,$j$. In a nutshell, partitioned graphs, as the one shown in Fig.~\ref{2}, are graphical constructions that group together in a \textit{node} the \textit{fg-sites} that fulfil the same connection properties (see the conditions listed above).

\begin{figure}[ht!]
    \centering
    \begin{subfigure}[p]{0.4\textwidth}
    \caption{}
         \includegraphics[width=\textwidth]{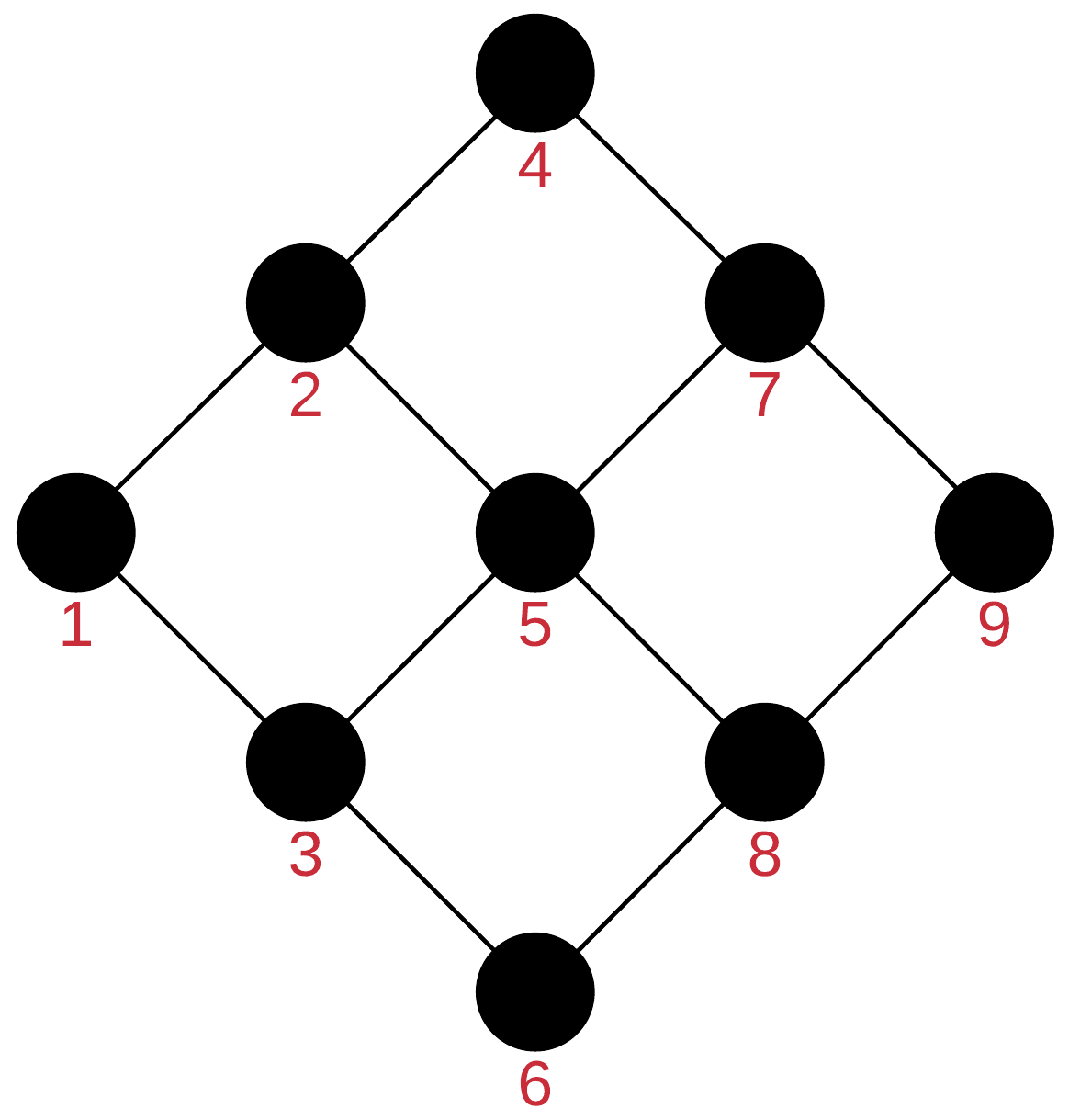}
        \label{a4full}
    \end{subfigure}
     ~
    \begin{subfigure}[p]{0.48\textwidth}
    \caption{}
        \includegraphics[width=\textwidth]{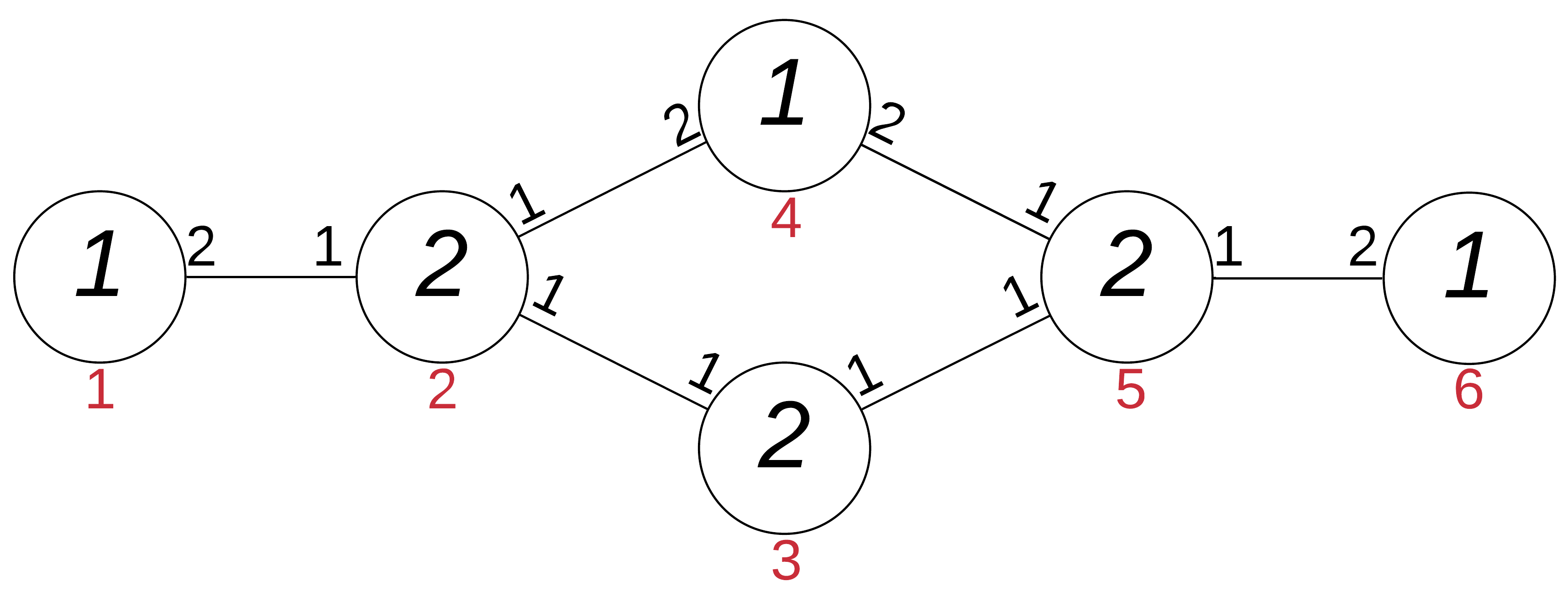}
        \label{a4a}
    \end{subfigure}
    ~ 
    \begin{subfigure}[p]{0.48\textwidth}
    \caption{}
        \includegraphics[width=\textwidth]{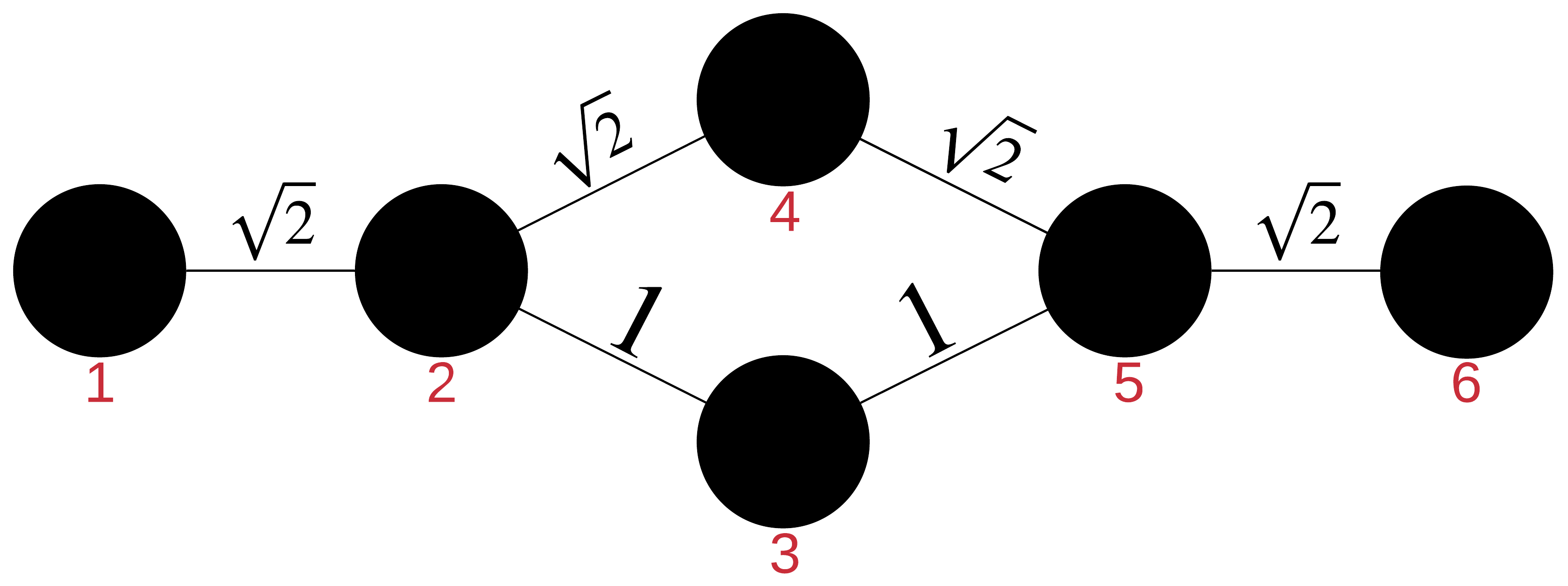}
        \label{a4b}
    \end{subfigure}
    \caption{The $3\times3$ square graph together with the corresponding partitioned graph and the quotient graph. The numbers \new{drawn on top of the edges} in the partitioned graph describe the coupling degrees of each node, \new{$d_i$}, and the numbers on top of the edges in the quotient graph denote the \new{effective} coupling strengths, $J_{i,j}$. Note that for the initial full $3\times3$ square graph all coupling strengths are $J_{i,j}=1$. The red numbers at the bottom of each circle are labelling the sites of the graphs (a) and (c) and the nodes of the partitioned graph (b).}\label{a4}
\end{figure}

Once the partitioned graph is obtained, the quotient graph can be easily defined. The locations of the qubits in the quotient graph are determined by its site structure, and this site structure is identical to the node structure of the partitioned graph. We will refer to the qubit sites of the quotient graph as \textit{qg-sites}, to distinguish these from the qubit locations in the full graph which are determined by the \textit{fg-sites}. \new{While in the full graph all the \textit{fg-sites} were uniformly coupled with $J=1$, in the quotient graph each \textit{qg-site} is now interacting with the adjacent \textit{qg-sites} through an effective coupling strength $J_{1,2}=\sqrt{d_1d_2}$. The quotient graph and the full graph present the same dynamics and therefore the quantum transfer abilities of both graphs are identical.} 
%Old sentence: The quotient graph presents the same dynamics, and therefore quantum transfer abilities, as the original non-partitioned graph, in which all the \new{sites} were uniformly coupled with $J=1$.

\begin{figure}[h]
    \centering
    \includegraphics[width=0.48\textwidth]{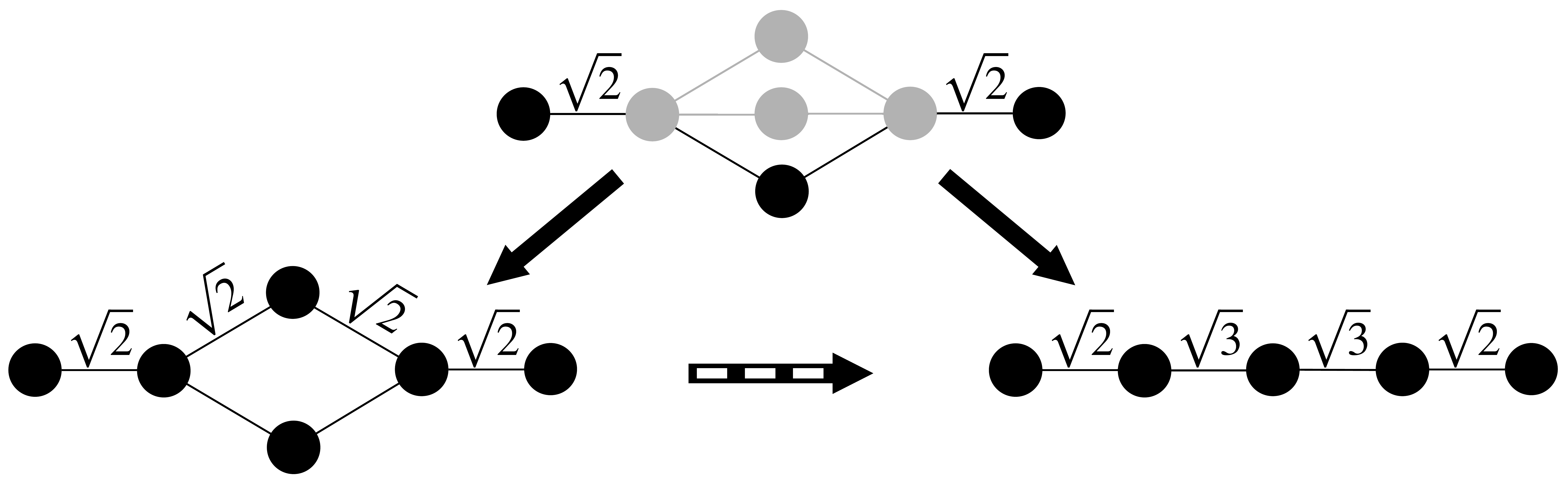}
    \caption{Sketch of the lift-and-quotient reduction \cite{Bachmann2012}.}
    \label{lift}
\end{figure}

We now move to the network arrangement considered in this work. First let us introduce the basic unit of our system: a single $3\times3$ square graph, as shown in Fig.~\ref{a4full}. \new{In \cite{chris2005}, Christandl et al. showed that this structure has perfect state transfer (PST) properties. The PST property entails that an excitation injected at the first \textit{fg-site}  is perfectly transferred to its mirror position (here, \textit{fg-site} 9) by the natural dynamics of the system. Here, by 'injected excitation' we mean the creation of a spin up state, $\vert \uparrow\rangle$, in a system that has all spins down, $\vert \downarrow\downarrow \ldots \downarrow\rangle$.
%The original \new{$P_3$ hypercube} together with the partitioned and quotient graph are represented in Fig.~\ref{a4}.
The complexity of the $3\times3$  structure} can be reduced by applying the aforementioned process of graph partitioning as shown in Fig.~\ref{a4}b, c. We note that for the presented quotient graph, Bachmann et al. introduced an additional lift-and-quotient reduction \cite{Bachmann2012} allowing for a further simplification of the graph to a linear chain, as shown in Fig.~\ref{lift}. Therefore, we do have some freedom in the way we perform that partition, as partitioning only the grey coloured sub-graph at the top of Fig.~\ref{lift} results in the bottom left graph, and if we partition everything, but the edges we get the bottom right graph. The bottom right graph is a linear chain, for which the qubit sites will be referred to as \textit{lc-sites}. \new{Importantly, when the full graph is  initialised in a normalised equal superposition between the \textit{fg-sites} that correspond to the initially excited \textit{qg-sites} of the quotient graph and, in turn, to the initially excited \textit{lc-sites} of the quotient linear chain,  the three graphs present the same dynamics.}

\begin{figure}[h]
        \centering
        \includegraphics[width=0.48\textwidth]{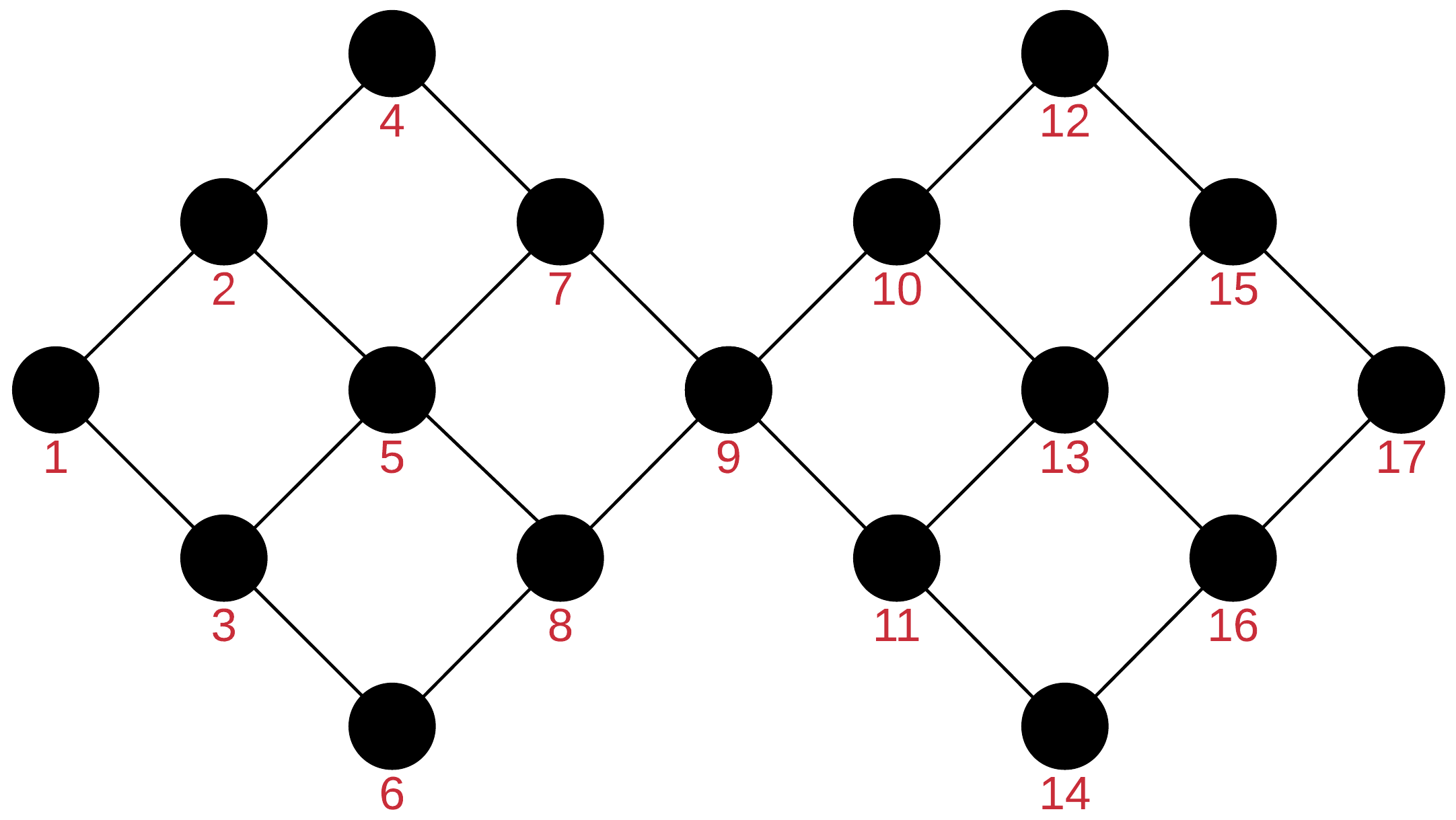}
        \caption{Graph of two interconnected $3\times3$ square graphs.}
        \label{twohypercubes}
\end{figure}

\begin{figure}[h]
    \centering
    \includegraphics[width=0.48\textwidth]{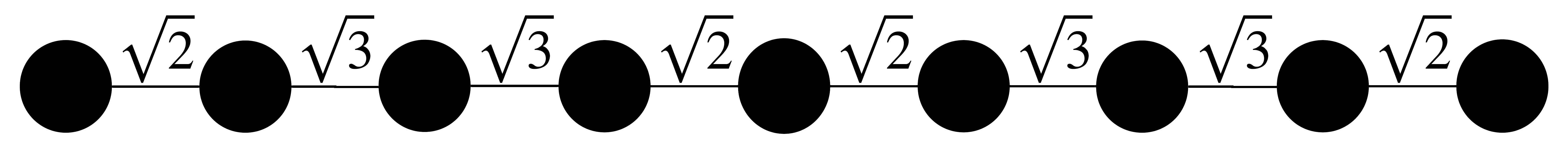}
    \caption{Quotient linear chain for the two interconnected $3\times3$ square graphs from Fig.~\ref{twohypercubes}.} 
    \label{chain}
\end{figure}

Given that the $3 \times 3$ square graph can be collapsed to a linear chain (see Fig.~\ref{lift}) \new{whilst} ensuring PST properties, we have chosen it as our basic unit to build a more complex system. For our entanglement generation protocol we interconnect two of these units (as shown in Fig.~\ref{twohypercubes}). \new{If we apply the partitioning method and the lift-and-quotient reduction to the two coupled $3\times3$ square graphs, we obtain a linear chain, as shown in Fig.~\ref{chain}. This can in turn be approximated to a trimer, a well-known structure capable of generating entanglement, as we will introduce in Sect.~\ref{subsec:abc}.} %We shall call the \new{sites} of this quotient linear chain (Fig.~\ref{chain}) \textit{lc-sites}.

\subsection{\label{sec:Unitary}Unitary Transformation Perspective}
An alternative and more physical perspective on the graph partition and quotient combined operation is to consider this operation as due to a unitary transformation. The reduction of the complexity of a graph to a simpler graph (with fewer coupled sites), or even a simple chain (with still fewer coupled sites), can be viewed as due to a unitary transformation. As we are considering the single excitation subspace of our system, the Hamiltonian has the same dimensionality as the site basis and the transformation redefines the definitions of (some of) the sites to superpositions of the original site basis. There are two criteria for the transformation. First, it should decouple some of the sites, to simplify the graph. Second, it should leave alone the definitions of the sites between which we seek identical dynamics in the reduced graph.

Clearly this perspective also works in reverse, in the sense that we could start with a simple graph or chain, and augment this with some additional uncoupled sites (which could be at zero energy or nonzero energy, dependent upon the form of the more complicated graph sought). Then, a unitary transformation can be chosen to redefine the site basis and involve the uncoupled sites in a more complex graph. If in this reverse approach the objective is again a network with identical dynamics between certain sites, these sites should be invariant under the transformation. 

In both cases (graph simplification and graph expansion) where the sites of interest for the dynamics are not invariant under the transformation, there are clearly still equivalent dynamics in the two graphs. However these will involve site superposition states, as related by the transformation.

We will refer to this unitary perspective in relation to the specific examples discussed in this paper. 

\subsection{\label{subsec:abc}ABC Configuration}
\begin{figure}[h!]
    \centering
    \includegraphics[width=0.48\textwidth]{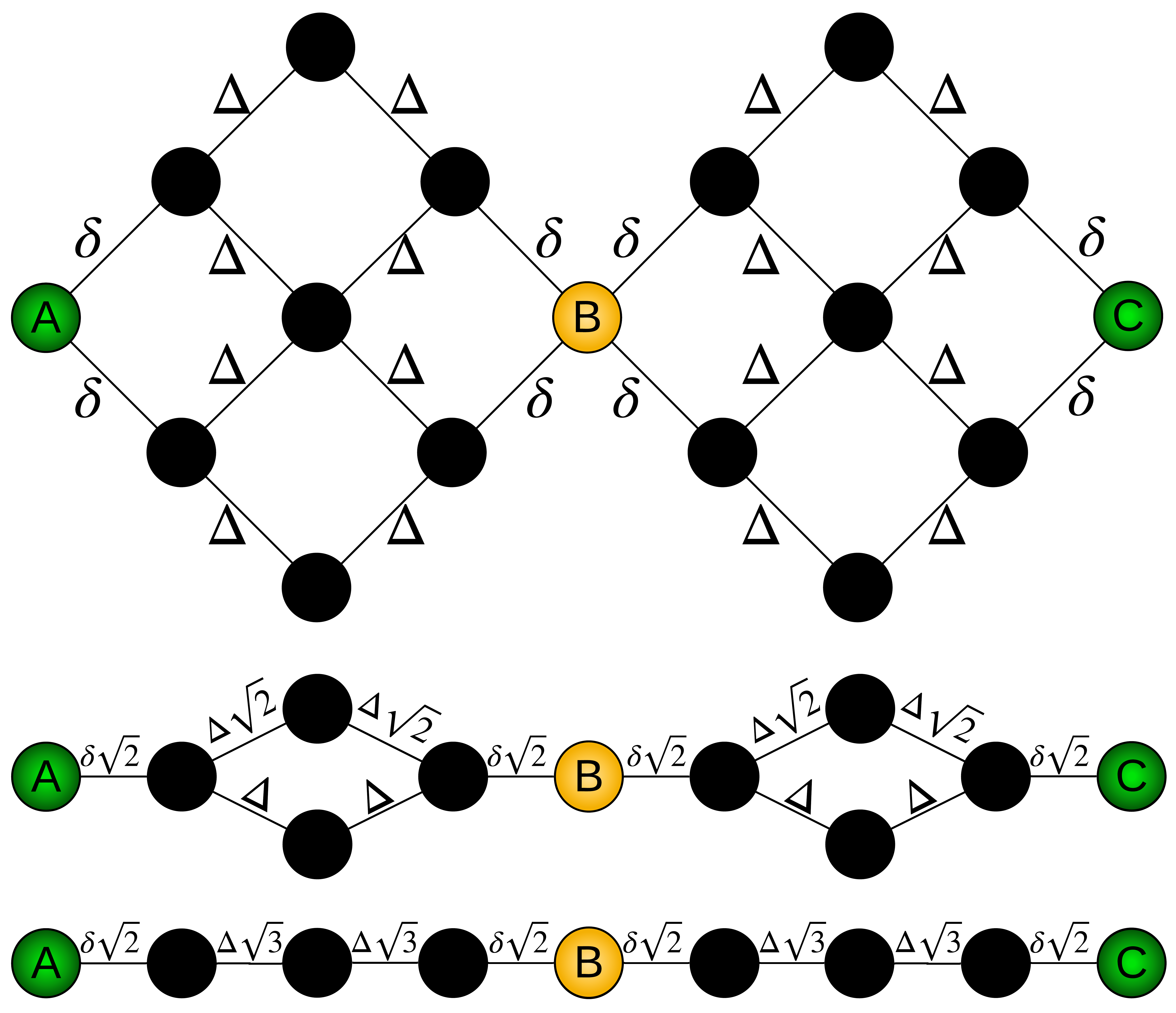}
    \caption{ABC configuration of the full two interconnected $3\times3$ square graphs (top), the quotient graph (middle) and the quotient linear chain (bottom).} 
    \label{ABC}
\end{figure}

%\green{The entanglement properties of graphs depend on their couplings. Hence, one could try to enlarge the entanglement of a graph structure by adjusting the coupling strengths. However, a graph where each coupling has a different strength is not the most practical for realisations. We therefore use a coupling configuration of only two parameters, called ABC configuration \cite{Estarellas2017}.} MARTA: THIS IS NOT NEEDED
We now extend what in \cite{Estarellas2017} is called ABC configuration to our specific graph structure. This configuration is attained by imposing a coupling distribution of two different energies, $\Delta$ and $\delta$, that results in having three sites (named A, B and C) distributed symmetrically and weakly coupled ($\delta$) to the rest of the system, such that they appear to be defects in an otherwise strongly coupled ($\Delta$) graph. The reason that makes this particular configuration interesting is that it can be approximated to a tunable trimer chain, which has the ability to dynamically create a maximally entangled Bell state between the edge sites when the system is initialised \new{by injecting an excitation} in the middle site \cite{Estarellas2017,Wilkinson2018}. Figure~\ref{ABC} shows the result of applying this configuration in our full graph and its two quotient structures.

From the unitary transformation perspective, the ABC system has seventeen single-excitation eigenstates, with eigenenergies as given in Fig.~\ref{EnergyTable}. The first partition and quotient operation is effected by a transformation that acts only on the black sites in Fig.~\ref{ABC} and leaves A, B and C invariant. This decouples six sites (two at positive energy $\sqrt{2} \Delta$, two at equal magnitude negative energy $-\sqrt{2} \Delta$ and two at zero energy), clearly without changing the overall spectrum of the system. The resultant non-trivial network is the 11-site quotient graph shown. A further unitary transformation, again acting on only the black sites and leaving A, B and C invariant, can decouple a further two zero energy sites to leave the final 9-site chain, with the spectrum given in Fig.~\ref{EnergyTable}.
\begin{figure}
\begin{center}
\begin{tabular}{ |l|c| } 
 \hline
  & \\
 Six eigenenergies that & $\sqrt{2} \Delta$ , $\sqrt{2} \Delta$ \\ 
 decouple in the reduction & 
 $-\sqrt{2} \Delta$ , $-\sqrt{2} \Delta$ \\ 
 from 17-site to 11-site & 0 , 0 \\ 
  & \\
 \hline
   & \\
 Two eigenenergies that &  \\
 decouple in the reduction & 0 , 0 \\ 
 from 11-site to 9-site &  \\ 
   & \\
 \hline
  & \\
 Nine eigenenergies that & $\pm \sqrt{3\delta^2+3\Delta^2 \pm\sqrt{\delta^4+9\Delta^4}}$ \\
 remain in the 9-site & $\pm \sqrt{\delta^2+3\Delta^2 \pm\sqrt{\delta^4+9\Delta^4}}$ \\ 
 quotient linear chain & 0 \\ 
  & \\
 \hline
\end{tabular}
\end{center}
 \caption{The seventeen single-excitation eigenenergies of
 the network shown in Fig.~\ref{ABC}, identifying those that decouple with the reductions.}
 \label{EnergyTable}
 \end{figure}

\subsection{\label{sec:mes} Entanglement-of-formation as a measure of graph performance}

To determine the effectiveness of any particular spin chain or graph to generate entangled states, a quantitative measurement of bipartite entanglement is needed. For this, we will use the entanglement of formation (EOF) \cite{wootters2001}. The EOF between qubits $A$, $C$ is defined by,
\begin{equation}
    \mathrm{EOF}_{AC}=-x\log_{2}x-(1-x)\log_{2}(1-x),
\end{equation}
where $x=\frac{1+\sqrt{1-\tau}}{2}$ and $\tau=(\text{max}\{0,\lambda_1-\lambda_2-\lambda_3-\lambda_4\})^2$. $\lambda_i$ is the square root of the $i^{th}$ eigenvalue of the matrix $\rho_{AC}\widetilde{\rho_{AC}}=\rho_{AC}[(\sigma^{A}_{y}\otimes\sigma^{C}_{y})\rho^*_{AC}(\sigma^{A}_{y}\otimes\sigma^{C}_{y})]$, ordered such that $\lambda_1>\lambda_2>\lambda_3>\lambda_4$. $\rho_{AC}$ is the reduced density matrix for sites A and C that result from tracing out the rest of the system, such that $\rho_{AC}=tr_\mathrm{rest}(\rho)$.

The EOF ranges between $0$ and $1$, with $EOF=1$ indicating that the state comprising two qubits is maximally entangled. \new{In our work, the EOF magnitude allows us to quantitatively assess how good is the entangled state generated from the natural dynamics of the graph. The EOF has also been used to assess the quality of other entangled state generation protocols in spin chains \cite{Estarellas2017,Wilkinson2018}}

\section{\label{sec:res}Results}

In order to study the dynamics of the system, we solve the time-independent Schrödinger equation through exact diagonalisation of the Hamiltonian matrix \cite{estarellas2018}. Note that, as already mentioned, the three structures (full graph, quotient graph and quotient linear chain) will have the same dynamics for injection and extraction at sites A, B and C. For that, we initialise the system to a spin up, $\vert 1\rangle$, at site B and all spins down, $\vert0\rangle$, in the rest of the graph. We then let the state evolve through its natural dynamics and calculate the EOF versus time. Because we based our structure on the trimer chain, the dynamics of the EOF will look like a Rabi oscillation which corresponds to the entangling and disentangling of the state comprising sites A and C. \new{In the remaining of this section, we study how the amplitude and period of such oscillations depend on the chosen coupling ratio, $\frac{\delta}{\Delta}$, and how such ratio affects the time one needs to wait to obtain the maximum EOF peak.} We also investigate how the presence of random fabrication errors (diagonal and off-diagonal disorder) affects differently the dynamics of the three graph structures, giving different results in terms of robustness.
\new{We will use natural units, so that $\hbar=1$.}

\subsection{\label{sec:ent}Entanglement generation}

\begin{figure}[h!]
    \centering
    \begin{subfigure}[p]{0.48\textwidth}
        \includegraphics[width=\textwidth]{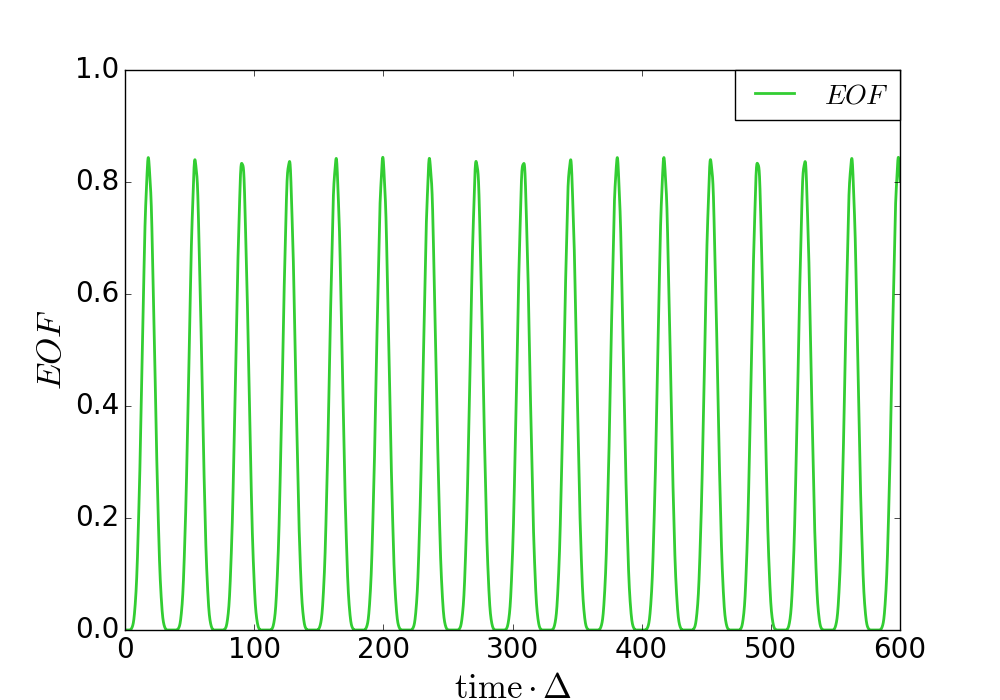}
    \label{12a}
    \end{subfigure}
    \begin{subfigure}[p]{0.48\textwidth}
        \includegraphics[width=\textwidth]{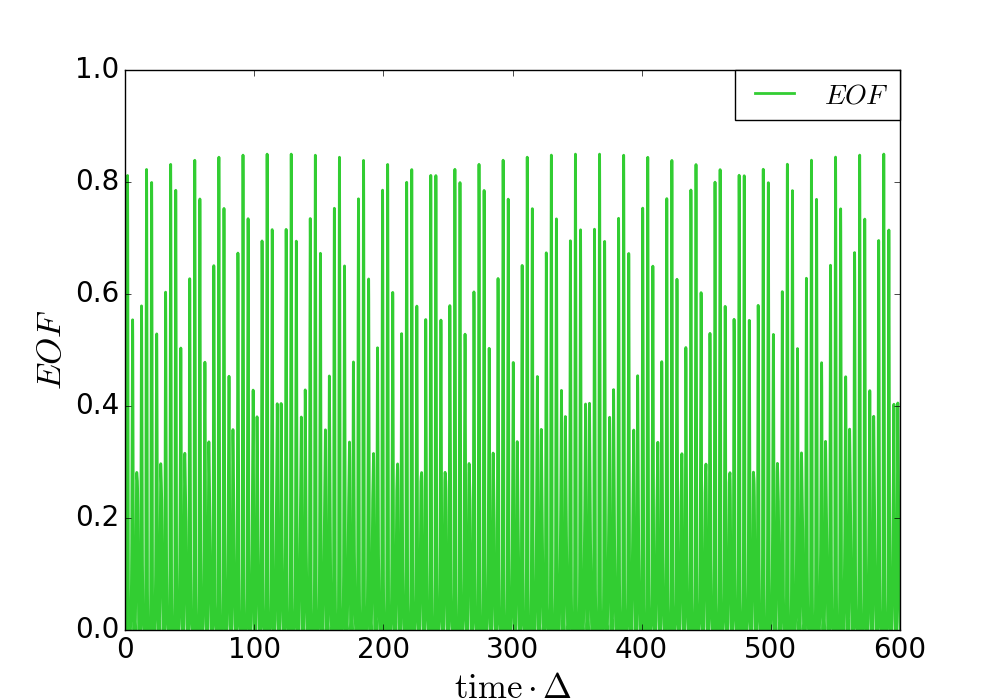}
        \label{12b}
    \end{subfigure}
    \caption{Dynamics of the EOF between sites A and C \new{for the  coupling ratios} $\frac{\delta}{\Delta}=\SI{0,1}{}$ (top) and  $\frac{\delta}{\Delta}=1$ (bottom). A coupling ratio of $\frac{\delta}{\Delta}=1$ corresponds to the case with uniform couplings.}
    \label{12}
\end{figure}

In Fig.~\ref{12}, we show the EOF dynamics for two different coupling ratios, $\frac{\delta}{\Delta}= \SI{0,1}{}$ and  $\frac{\delta}{\Delta}=1$. The peaks for each of the two scenarios present different periodicity and \new{related} amplitudes. In Fig.~\ref{12}, it is also apparent that the larger the coupling ratio, the faster the oscillations, meaning that the entangled state is generated earlier. For $\frac{\delta}{\Delta}=1$, the first EOF peak happens at $t_1\cdot \Delta=\SI{1,97}{}$ and for $\frac{\delta}{\Delta}= \SI{0,1}{}$, \new{at  $t_1\cdot \Delta=\SI{18,02}{}$. }

\subsubsection{EOF at the first peak}

Our first approach to compare the effect of the different coupling ratios in our protocol is to investigate the first EOF peak. From an experimental point of view, the evaluation of the first peak is a useful metric as it will be most likely to fall ahead of the decoherence times of the experimental realisation. The dependence of the time when the first entanglement operation happens can be analytically approximated from the reduced trimer as done in \cite{Wilkinson2018}. From that, we obtain
\begin{equation}
    t_P\cdot \Delta=\frac{\pi}{\sqrt{3+\left(\frac{\delta}{\Delta}\right)^2-\sqrt{9+\left(\frac{\delta}{\Delta}\right)^4}}} 
\end{equation}
as the estimate of the period of the EOF oscillation and of the time needed for an excitation injected at site B to propagate to the edges and return to its initial state. Thus, the entangled state will be formed for the first time at approximately $t_P\cdot \Delta/2$. \new{ Also, in our calculations we have numerically obtained the exact time, $t_1$, at which the first EOF peaks occurs within a time window equal to $t_P$.}

\begin{figure*}[ht!]
\centering
\resizebox{0.85\textwidth}{!}{
  \includegraphics{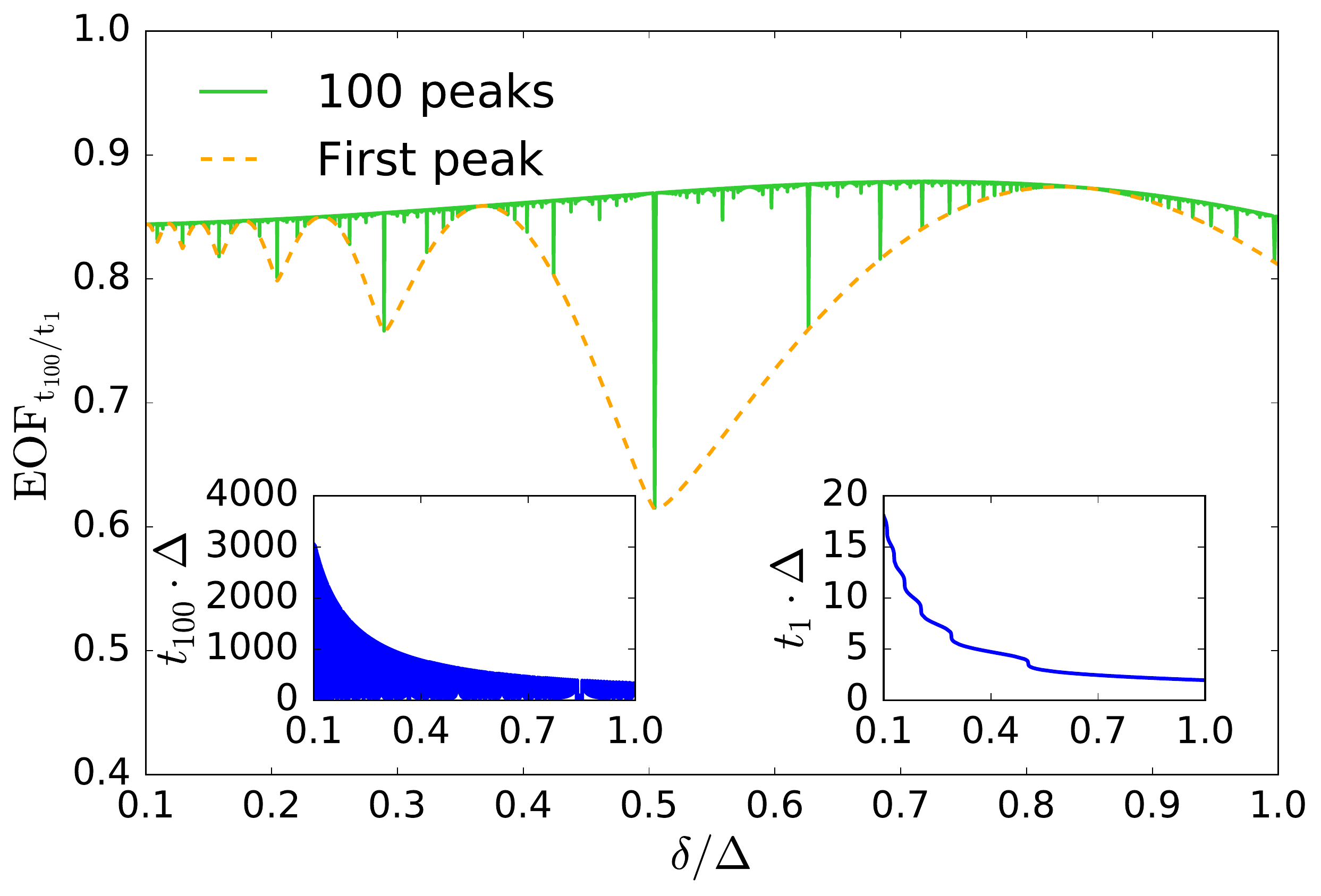}}
\caption{Orange dashed curve: EOF at $t_1$ plotted against the coupling ratio $\frac{\delta}{\Delta}$. This value is obtained as the highest entanglement found in a time window of $t_P$. Green solid curve: maximum EOF in a time window of $100 \cdot t_P$, \new{$\mathrm{EOF}_{t_{100}}$}, against the coupling ratio $\frac{\delta}{\Delta}$. In the left inset, the time $t_{100}$ when the maximal entanglement within $100\cdot t_P$ occurs is plotted against the coupling ratio. In the right inset, the time $t_1$ when the first entanglement peak occurs is plotted against the coupling ratio.}
\label{13}       
\end{figure*}

\begin{figure}[h!]
    \centering
    \resizebox{0.48\textwidth}{!}{
        \includegraphics{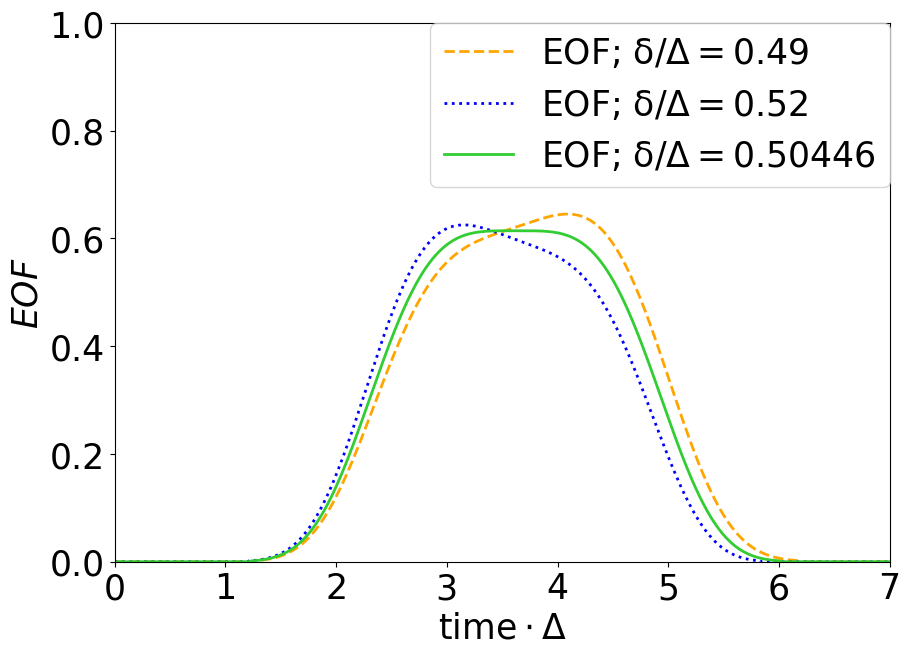}}
    \caption{EOF for ratios around the orange dashed curve minimum \new{at} $\frac{\delta}{\Delta}=0.50446$ in Fig.~\ref{13}. The shifting of the maximum from the right hand side of the peak to the left hand side explains the staircase like profile of the left inset in Fig.~\ref{13}. Orange dashed: EOF vs. time for a coupling ratio slightly smaller than the rightmost minimum of the orange dashed curve in Fig.~\ref{13}. Green solid: EOF vs. time for a coupling ratio equal to the rightmost minimum of the orange dashed curve in Fig.~\ref{13}. Blue dotted: EOF vs. time for a coupling ratio slightly larger than the rightmost minimum of the orange dashed curve in Fig.~\ref{13}.}
    \label{t1explanation}
\end{figure}

\new{The coupling ratio dependence of the first EOF peak for the two interconnected $3\times3$ square graphs is shown from the orange (dashed) curve of Fig.~\ref{13}. This dependence is identical for the three structures (full graph, quotient graph and quotient chain). This curve shows an oscillatory behaviour, with amplitude increasing and frequency decreasing with the coupling ratio.} We observe that the highest EOF attained is $\mathrm{EOF}=\SI{0.8745}{}$ for a coupling ratio of $\frac{\delta}{\Delta}=\SI{0.82846}{}$. In the right inset of Fig.~\ref{13} we show how the time $t_1$ decays as the coupling ratio increases, a result that is in agreement with \new{its analytical approximation}. The dependence of $t_1$ on the coupling ratio has a staircase-like profile; the quick vertical drops occur at the ratios corresponding to the minima of the orange dashed curve in the main panel. This behaviour can be understood if we look at a few consecutive slices of the dynamics for the region close to those minima. Figure~\ref{t1explanation} shows three nearby points to the minimum close to $\frac{\delta}{\Delta}=0.5$. From the lower $\frac{\delta}{\Delta}$ to the higher, we observe how the EOF curve goes from having a clear maximum to reach a plateau, and then the maximum can be distinguished again. This transition results in a $t_1$ step (note how the maximum that was initially at the right-hand side appears at the left side after reaching the flat plateau), as seen in the right inset in Fig.~\ref{13}. This behaviour can be observed at all of the minima of the orange dashed curve in  Fig.~\ref{13}.

\subsubsection{Comparison to `normal' linear chain}
\new{When changing the coupling ratio of the quotient linear chain in Fig.~\ref{ABC} from $\sqrt{2}\delta/\sqrt{3}\Delta$ to $\delta/\Delta$ we find that the dynamics is the same, but with a rescaled factor $\sqrt{3}/\sqrt{2}$. In particular, the first EOF peak is reached later, at a time longer by $\sqrt{3}/\sqrt{2}$.
So if one has only two specific couplings $\delta$ and $\Delta$ available due to experimental constraints, then using the full graph generates faster dynamics (a shorter $t_1$) than a coupled linear chain generated by those same two couplings. In this sense, the dynamics of the full graph is equivalent to the dynamics of an `enhanced' spin chain with coupling ratio increased by a factor $\sqrt{3}/\sqrt{2}$.}

\subsubsection{\label{sec:rob}Entanglement within a longer time}
            
A different approach to compare the coupling ratio dependence is to use the maximum entanglement generated over a larger time window. \new{We denote $t_{100}$ the time at which the highest entanglement $\mathrm{EOF}_{t_{100}}$ occurs in a time window of 100 $t_p$ periods (note that $t_P$ will depend on the coupling ratio)}. Figure~\ref{13} shows the dependence of \new{$\mathrm{EOF}_{t_{100}}$} (green solid profile) with the coupling ratio. For this scenario and a coupling ratio of $\delta/\Delta=\SI{0,72018}{}$ we get the highest maximum entanglement, \new{$\mathrm{EOF}_{t_{100}}=\SI{0,8787}{}$}. The left-hand inset in Fig.~\ref{13} shows the time $t_{100}$ when the maximal entanglement occurs. The upper limit is given by the time window $t=100\cdot t_P$ and is decreasing with the coupling ratio. We will later have a more detailed look at this time behaviour.

The green solid line in Fig.~\ref{13} outlines an upper limit with respect the orange dashed curve, but it also displays a few downward outliers. For such cases, the highest possible entanglement occurs at a time out of the selected time frame due to the presence of secondary oscillations of a large period (examples of secondary oscillations can be seen in Fig.~\ref{12}). We will call these outliers `downwards peaks'. Note that the `downwards peaks' are very sharp so we need a high precision in the coupling ratio to identify them. The sharpness of the `downwards peaks' also underlines the quick change in the secondary oscillations just by slightly modifying the coupling ratio.

\subsubsection{Ratios for perfect periodicity}
In Fig.~\ref{13}, for certain ratios, the green solid curve shows `downwards peaks' touching the orange dashed curve; this behaviour implies that that no EOF peak is higher than the first peak in the specified time frame of $100\cdot t_P$. However, many of these downward peaks remain when considering an arbitrarily large time window, suggesting that no EOF peak is higher than the first peak. This either means that the first EOF peak is \new{the} highest or that all EOF peaks \new{have the same height (periodic system). We shall see first through inspection of one of these `downward peaks' and then analytically that the latter is true}.

\begin{figure}[h!]
    \centering
    \resizebox{0.48\textwidth}{!}{
        \includegraphics{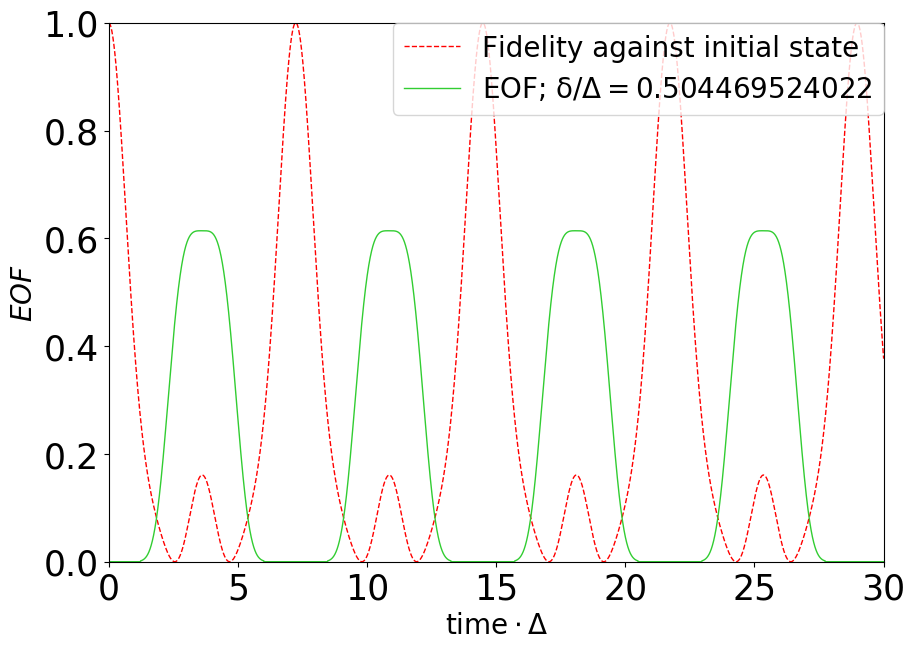}}
    \caption{\new{Green solid: EOF at the coupling ratio of one of the `downwards peaks' (corresponding to the rightmost minimum of the orange dashed curve in Fig.~\ref{13}). Red dashed: fidelity against the initial state. EOF and fidelity show that the system returns periodically to its initial state.}}
    \label{integerConditionGraph}
\end{figure}

As shown in Fig.~\ref{integerConditionGraph}, the state of the system at those specific coupling ratios is fully periodic. \new{The system returns exactly to its initial state, and the second and consecutive EOF peaks have the same shape as the first}. We conclude that all the peaks are the same for the specific coupling ratios showing `downwards peaks'. 

There are multiple \new{coupling} ratios for which this is true, \new{and} the reason for this, \new{as well as} the precise values of these ratios, can be discovered through the analysis of the fidelity against the initial state.
For an initial state $\ket{\psi_0}$, the fidelity against a state $\ket{\psi_f}$, is defined as
\begin{equation}
    \mathcal{F}(t)=|\bra{\psi_f}e^{-it\mathcal{H}}\ket{\psi_0}|^2
\end{equation}
where $\mathcal{H}$ is the time-independent system's Hamiltonian. By diagonalising $\mathcal{H}$ one can obtain its eigenvalues and eigenvectors $\{E_i\}$ and $\{\ket{\phi_i}\}$, and the fidelity can be written as
\begin{equation}
    \mathcal{F}(t)=\sum_{i,j}\alpha_{ij}e^{it(E_j-E_i)},
\end{equation}
where we have defined\\ $\alpha_{ij}=\braket{\phi_j|\psi_f}\braket{\psi_f|\phi_i}\braket{\phi_i|\psi_0}\braket{\psi_0|\phi_j}$. By noting that $\alpha_{ij}=\alpha_{ji}^*$, and $e^{it(E_j-E_i)}=(e^{it(E_i-E_j)})^*$, we can see that the imaginary part of the $i,j$ term cancels with the imaginary part of the $j,i$ term. Therefore, as the diagonal terms, $\alpha_{ii}e^{it(E_i-E_i)}$ are real, the fidelity can be written as a sum of cosines:
\begin{equation}
    \mathcal{F}(t)=\sum_{i,j}\alpha_{ij}\cos((E_j-E_i)t).
\end{equation}
\\
\new{The dynamics we are interested in is the same for the full, seventeen \textit{fg-sites} graph, the quotient graph and the quotient linear chain. We shall focus our analysis here on the simpler case of the quotient linear chain (with an initial injection in the centre \textit{lc-site} B).}
For this system, there are nine eigenvectors, five even under reflection about \textit{lc-site} B and four odd under this reflection.
There are thus five eigenvectors which are not orthogonal to the (even) initial \new{state. These eigenvectors  correspond to the eigenenergies (taken from Fig.~\ref{EnergyTable}) } 
\begin{align}
    \pm E &:= \pm \sqrt{3\delta^2+3\Delta^2+\sqrt{\delta^4+9\Delta^4}} \\
    \pm E' &:= \pm \sqrt{3\delta^2+3\Delta^2-\sqrt{\delta^4+9\Delta^4}} \\
    E_0 &:=  \; 0 \;.
\end{align}
As $\alpha_{ij}=0$ when either \new{ $\ket{\phi_i}$ or $\ket{\phi_j}$ } is orthogonal to the initial state $\ket{\psi_0}$, these are the only eigenvalues that affect the time dependence of $\mathcal{F}(t)$. \new{We need to consider all combinations of these five eigenvalues in the term $\cos((E_j-E_i)t)$. By noting that cosine is an even function, it can be shown that in order to equal unity (and therefore for the fidelity to return to its initial state) the following must be satisfied:}
%By considering all combinations of these five eigenvalues in the term $\cos((E_j-E_i)t)$, and noting that cosine is an even function, it can be shown that for all the cosines in the sum to equal unity (and therefore for the fidelity to return to its initial state) the following must be satisfied:

\begin{align*}
    Et &= 2\pi n_1 \\
    E't &= 2\pi n_2 \\
    2Et &= 2\pi n_3 \\ 
    2E't &= 2\pi n_4 \\ 
    2(E-E')t &= 2\pi n_5 \\
    2(E+E')t &= 2\pi n_6 \\
\end{align*}for some integers $n_i$. Of course, these equations are not independent, if the first two are satisfied, then so are the rest. Therefore, when $E'n_1=En_2$, $\mathcal{F}(t)=1$ at time $t=\frac{2\pi n_1}{E}=\frac{2\pi n_2}{E'}$. 
\\
Using this integer condition, and the formulae for the energies, we can derive a formula which will tell us, \new{for  given} $n_1, n_2$, what coupling ratio ensures that the condition $E'n_1=En_2$ is true.
\new{When this condition is satisfied for integers $n_1,n_2$, the system exhibits periodic dynamics. When $n_2=1$, the system returns to its initial state after a single peak in EOF, implying that every EOF peak will be the same height. For example, the dynamics shown in Fig.~\ref{integerConditionGraph}, for a ratio $\frac{\delta}{\Delta}=0.504469524022$, is produced when $n_2=1$ and $n_1=3$.}

\subsubsection{Time Behaviour around `Flat Coupling Ratios'}

\begin{figure}[h!]
\centering
\resizebox{0.48\textwidth}{!}{
  \includegraphics{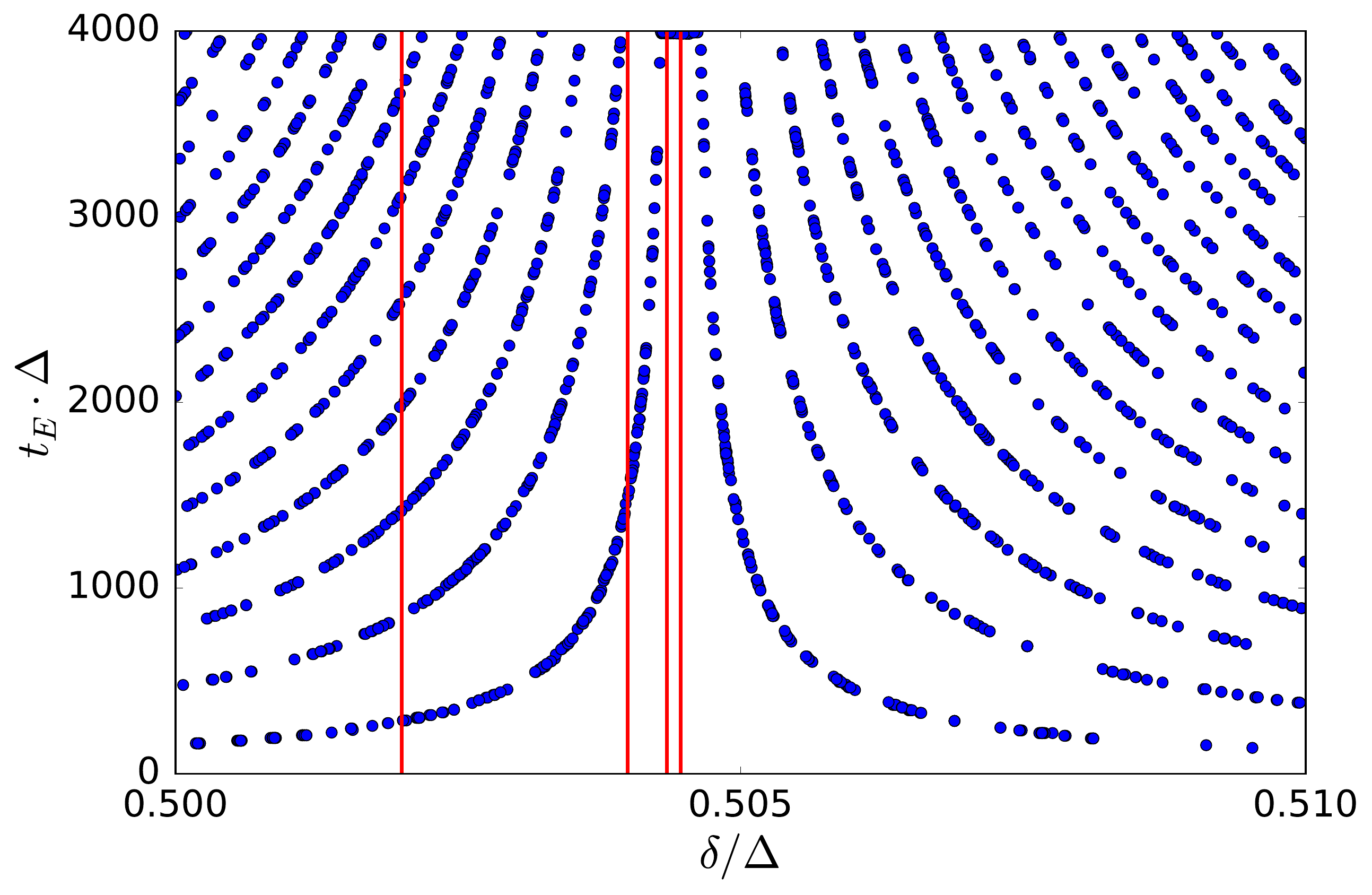}}
\caption{Time behaviour around the coupling ratio \new{corresponding to the absolute} minimum of the orange dashed curve in Fig.~\ref{13}. Every point stands for the time \new{$t_E$} when the highest EOF is observed in a time window of $t\cdot \Delta=4000$. To be sure that the overall shape of the plot was not due to the sampling of the coupling ratios, we have chosen 2000 random coupling ratios in the range [\SI{0.50}{},\SI{0.51}{}].}
\label{time4000}       
\end{figure}

Here, we consider coupling ratios within a small interval around the \new{minima} of the orange dashed curve of Fig.~\ref{13} that \new{coincide with the downwards peaks of the green} solid curve of Fig.~\ref{13}. We investigate the time when the highest EOF occurs by considering a fixed time window. In Fig.~\ref{time4000}, we show the highest EOF \new{within $0\leq t\cdot \Delta \leq 4000$} for small variations of the coupling ratios around $\delta/\Delta=0.505$. The period of the secondary oscillations becomes longer if we get closer to the minimum of the orange dashed curve in \new{ Fig.~\ref{13}, therefore,} the number of the EOF maxima from the secondary oscillations within the observed time window decreases. \new{This behaviour is confirmed in Fig.~\ref{time4000}. Here, each of the blue curves tracks the evolution of one EOF maxima as the coupling ratio $\frac{\delta}{\Delta}$ varies and $0\leq t\cdot \Delta \leq 4000$. By considering vertical cuts at specific coupling ratios (the red lines), we can see that the number of blue curves intersecting each cut decreases as we get closer to the coupling ratio $\frac{\delta}{\Delta}=0.504469524022$ (minimum of the orange curve in Fig .~\ref{13}).} \new{More details are in Appendix where we show the dynamics of the EOF for the specific coupling ratios corresponding to the vertical lines in Fig.~\ref{time4000}. Close to this coupling ratio, the highest EOF within the chosen time window occurs just before $t\cdot \Delta=4000$,  but the absolute highest EOF (a maxima of the secondary oscillation) occurs outside the time window.} As the coupling ratio approaches the minimum of the orange dashed curve in Fig.~\ref{13} the time when the absolute highest EOF occurs goes to infinity. Then, we cannot observe any secondary oscillations. We call all the coupling ratios which lead to an infinite long period of the secondary oscillations `flat coupling ratios'. 

\subsection{\label{sec:loc} Stability against errors: Random Static Disorder}

\begin{figure*}[ht!]
\resizebox{\textwidth}{!}{
  \includegraphics{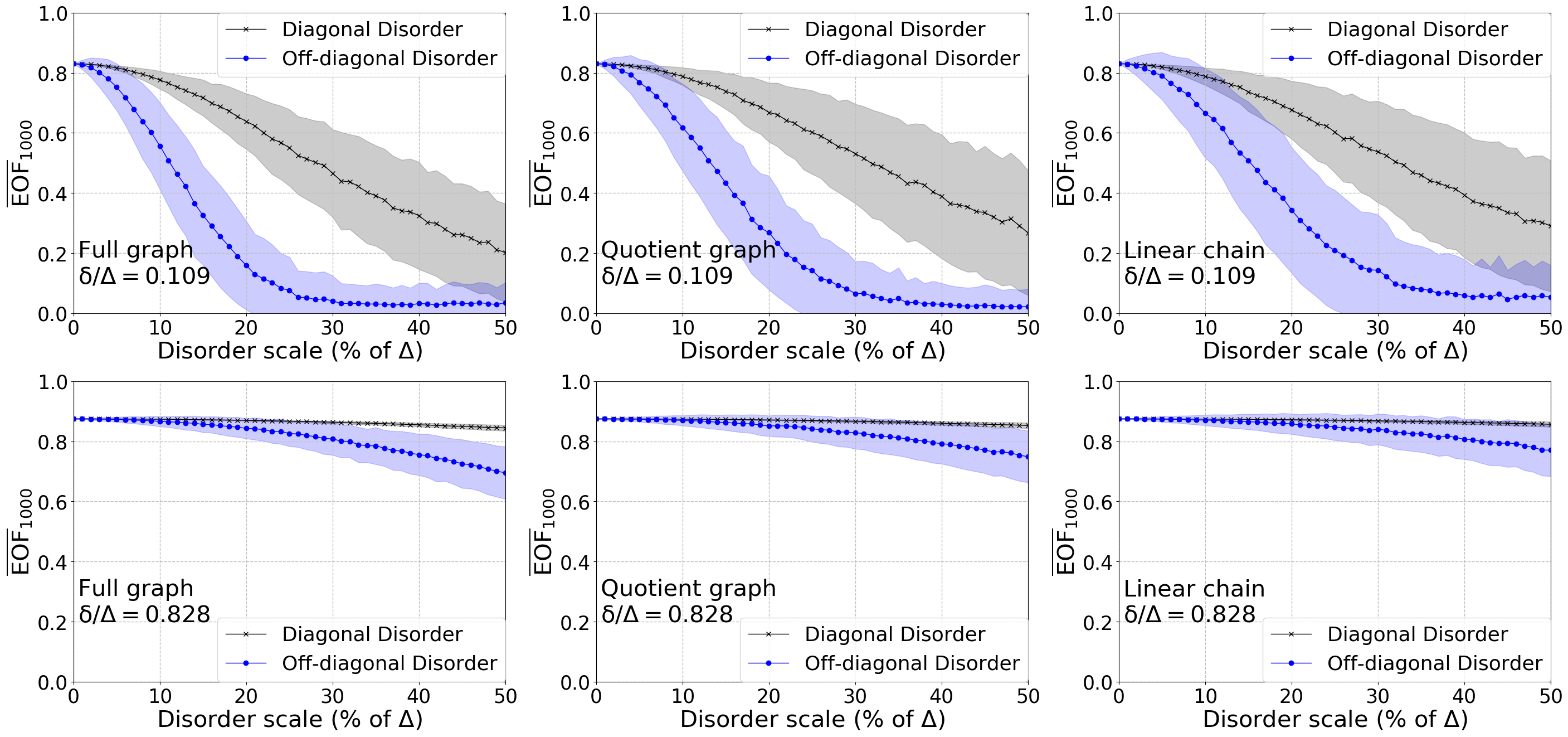}}
\caption{Stability against disorder. \textit{Top:} $\frac{\delta}{\Delta}=0.109$; \textit{Bottom:} $\frac{\delta}{\Delta}=0.828$. \textit{Left:} Full graph; \textit{Middle:} Quotient graph; \textit{Right:} linear chain. Off-diagonal disorder is shown in blue with dots, and diagonal disorder is shown in black with crosses. \new{The shaded areas represent the standard deviation}. The three graph structures display a very similar response to both disorders. }
\label{errorsResulsFull}       
\end{figure*}

So far we have discussed the behaviour of an ideal system, with no errors in the coupling strengths between sites and where all of the on-site energies are precisely equal and scaled to zero. This is of course an unrealistic situation and a consideration of how robust a system is to errors in manufacturing is critical to determining its practical feasibility. Following \cite{ronke2011_1}, to study our systems' robustness we introduce two types of static disorder. The first, \textit{diagonal disorder}, consists of adding random perturbations to the diagonal terms of the Hamiltonian, and represents random differences between the on-site energies of the qubits. \new{ The second type of disorder we apply, \textit{off-diagonal disorder}, represents random errors in the couplings between qubits. This is incorporated into the simulation by adding random perturbations to the nonzero off-diagonal terms of the Hamiltonian.} These two types of perturbation encompass a wide range of potential fabrication defects; however, we leave the study of other sources of errors or decoherence for a follow-up work. To see the effect of these errors on the dynamics of the system, we apply these perturbations to two different coupling scenarios: a coupling ratio of $\frac{\delta}{\Delta}=0.828$ (corresponding to the rightmost maximum of the orange dashed curve in Fig.~\ref{13}) and a coupling ratio of $\frac{\delta}{\Delta}=0.109$ (corresponding \new{to the left-most} minimum of the orange dashed curve in Fig.~\ref{13}). \new{We scale the errors via a disorder scale, $D$, ranging from $0\%$ to $50\%$ of the strong coupling, $\Delta$. $D$ is then included into the couplings or energies as a dimensionless parameter such that $\overline{\epsilon}_i=\epsilon_i+Dr_i\Delta$ and $\overline{J}_{i,i+1}=J_{i,i+1}+Dr_i\Delta$, where $r_i$ is a random number generated from a uniform distribution between $0$ and $1$}. For both types of added disorder, we perform $1000$ random realisations for each value of $D$, and for each realisation we obtain the EOF at the time of the first peak in the unperturbed system, $t_1$, and then calculate the average.

In Fig.~\ref{errorsResulsFull}, we compare the robustness of the graph structure of Fig.~\ref{twohypercubes} to its quotient graph and quotient linear chain. \new{We find that the three graph structures show a quick decay of the averaged EOF for the coupling ratio $\frac{\delta}{\Delta}=0.109$, \new{especially for the off-diagonal disorder} (top panel). However, for the case of the coupling ratio $\frac{\delta}{\Delta}=0.828$ (bottom panel) we observe an excellent robustness, especially in the case of diagonal disorder.} Hence, there is a strong dependence \new{of the robustness of our protocol on} the chosen coupling ratio. In addition to this, there is a significant difference between the effects of diagonal and off-diagonal disorder, with the latter being much more damaging for both coupling ratio scenarios. \new{While the three graph structures show similar robustness, the linear chain is slightly more robust than the quotient graph which in turn is slightly more robust than the full graph.} It is important to note that the off-diagonal disorder simulates errors in the coupling between sites and therefore high levels of this disorder significantly change the level of connectivity within the graph. \new{This has an impact on how the wavefunction evolves in space. For example, a negligible coupling can become much larger due to error, losing the required strong/weak coupling pattern of the system and, de facto, resulting in the opening of new channels for the wavefunction to diffuse.}

\section{\label{sec:concl}Conclusions}

Using a graph \new{formed by} two interconnected $3\times3$ square graphs, we have shown that one can efficiently generate bipartite entanglement by preparing an initial state with an excitation in the middle site of the graph. We engineered the graph couplings with a strong and weak coupling distribution such that we obtain an $ABC$ configuration that can be approximated to the trimer chain, known to generate Bell pairs. We analysed the spin dynamics dependent on the ratio of the weak and strong coupling and found specific coupling ratios where the entanglement shows a perfect periodic behaviour. This behaviour, however, can be rapidly \new{lost by} a slight change on the coupling ratio. In such cases, we encounter secondary oscillations causing the entanglement peaks to have a different height and therefore showing a different EOF.

We used graph partition theory to derive \new{two additional} related graphs and showed that the above findings are identical for the three graphs. In addition, we discussed the partitioning from the physical perspective of unitary transformations applied to redefine some of the graph sites. All three graphs depicted in Fig.~\ref{ABC} show the same dynamics after an excitation is injected in the middle. \new{Moreover, the three graphs show the same dynamics if the initial state of the full graph is a normalised superposition between the \textit{fg-sites} which correspond to the initially excited \textit{qg-sites} of the quotient graph and, in turn, the initially excited \textit{lc-sites} of the quotient linear chain.} This gives experimentalists flexibility in their system's topology; for example, for certain hardware, a full graph could be more favourable to implement than a linear chain, or it could offer additional functionalities. In addition, if the set of available couplings is limited, the full graph is advantageous as it corresponds to a spin chain with faster dynamics than a spin chain constructed from the same available couplings.

Finally, we considered the robustness of the three systems. We found that there is a significant dependence on the ratio $\frac{\delta}{\Delta}$, as the systems with a ratio $\frac{\delta}{\Delta}=0.828$ are significantly more robust than the ones with ratio $\frac{\delta}{\Delta}=0.109$. 
\new{These two ratio correspond to a maximum and a minimum, respectively, of the curve describing how the height of the first entanglement peak varies with the coupling ratio. This suggests that ratios corresponding to a maximum of this curve lead to more robust devices, but a systematic analysis should be conducted to support this conclusion, which is beyond the scope of this work.}
We also noted that errors affecting the coupling between sites (`off-diagonal disorder') are more damaging to the entanglement generation protocol than errors affecting the on-site energies of the sites (`diagonal disorder'). The three graph structures show similar robustness, although the linear chain is slightly more robust than the quotient graph which in turn is slightly more robust than the full graph. These results suggest that a physical realisation of the systems shown in Fig.~\ref{ABC} should aim for a ratio of $\frac{\delta}{\Delta}=0.828$, as it not only produces a high EOF in a shorter time, but is also extremely robust.

We conclude that graph structures with only two different couplings $\delta$ and $\Delta$ can be used to generate robust bipartite entanglement. While a linear chain with the same dynamics as the graph is slightly more robust to errors, a linear chain with the same coupling as the graph displays a slower dynamics. \new{We show that, depending on the experimental demands, one has the freedom to choose between graphs or chains for the design of different quantum technology applications.}
\\

\section*{Conflict of interest}
The authors declare that they have no conflict of interest.

\section*{Acknowledgements}
J. Riegelmeyer would like to thank J. Risske for helpful discussions about the code written to run the simulations.
%\end{acknowledgements}

\bibliography{papers_bib}
\vspace{9pt}

%\newpage

\section*{Appendix: EOF for coupling ratios close to $\frac{\delta}{\Delta}=0.504469524022$}

\begin{figure*}[h]
\resizebox{\textwidth}{!}{
  \includegraphics{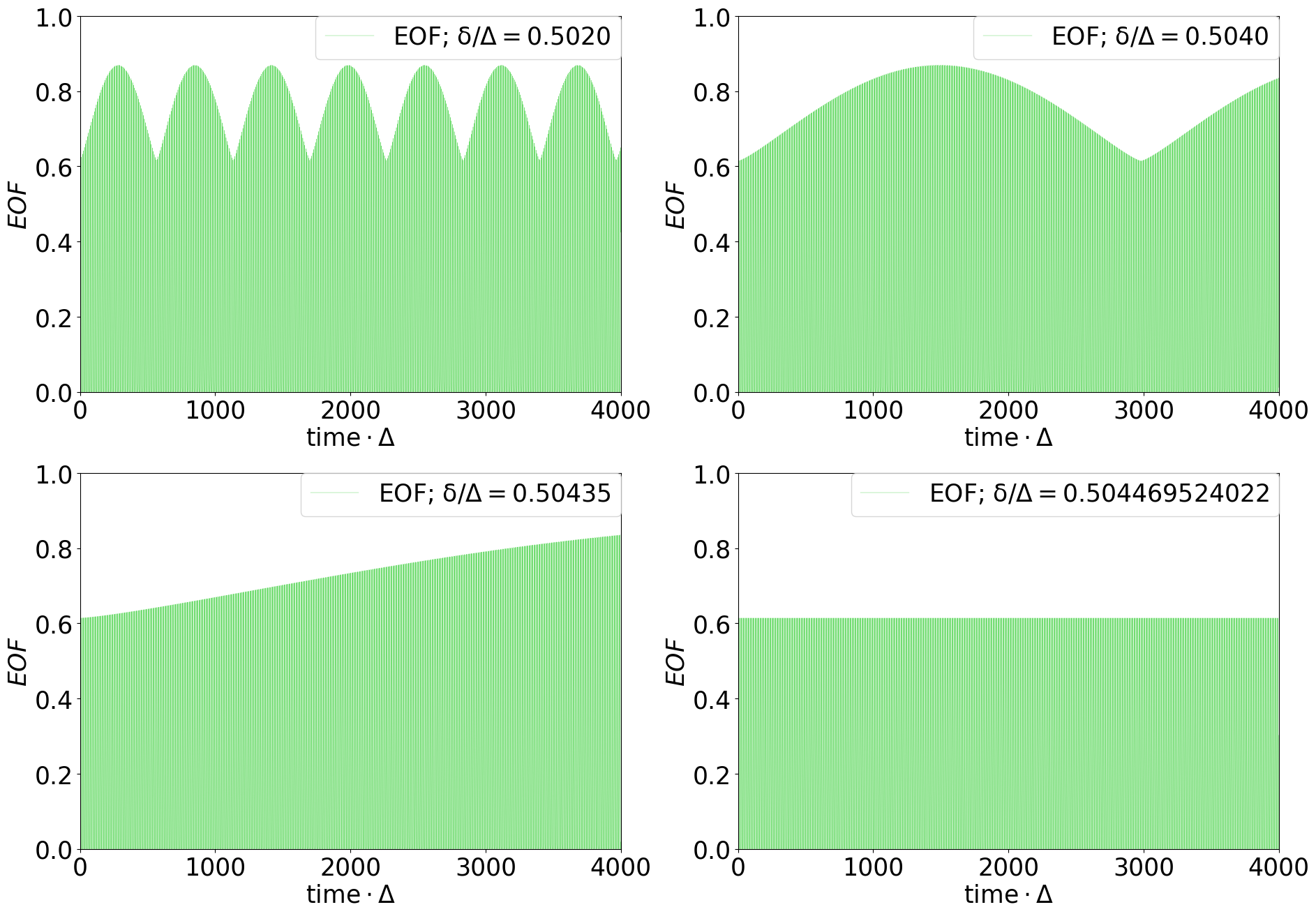}}
\caption{EOF vs time for coupling ratio corresponding to the vertical red lines shown in \new{Fig.~\ref{time4000}}. From top to bottom, left to right, the graphs show the EOF behaviour as the coupling ratio gets closer to the ``flat ratio" $\frac{\delta}{\Delta}=0.504469524022$. The top graphs show that for each coupling ratio there are multiple EOF maxima, and the bottom left graph shows that for ratios sufficiently close to the flat ratio, the first maximum will fall outside of the considered time window.}
\label{eofNearFlatRatio}       
\end{figure*}

In \new{Fig.~\ref{time4000},} the behaviour of the time of the \new{highest} EOF peak around a minimum of the orange dashed curve in Fig.~\ref{13} is shown. In this graph, the time, $t_E$, seems to reach a maximum of $t_E \cdot \Delta = 4000$ for ratios near to $\frac{\delta}{\Delta}=0.504469524022$; this is in fact a consequence of the \new{chosen simulation time window} cutting off at this time. As the bottom left graph in Fig.~\ref{eofNearFlatRatio} shows, systems with a coupling ratio in this region achieve their first maximum after \new{ $t \cdot \Delta = 4000$} (although the highest EOF in the considered time range is just before $t_E \cdot \Delta = 4000$). Figure \ref{eofNearFlatRatio} also offers an explanation for the multiple curves seen in \new{Fig.~\ref{time4000}}: as the system exhibits periodic behaviour, there are multiple EOF maxima for each coupling ratio. In \new{Fig.~\ref{time4000},} each curve intersecting with the red line corresponds to a different EOF maxima, which shift in time as the coupling ratio is changed. \new{Due to our simulation choosing only one EOF maximum for each coupling ratio, each coupling ratio is shown to have just one time value at which EOF is maximised, even when the dynamic is periodic.} This results in the \new{blue} curves in Fig.~\ref{time4000} not being continuous. However, in a true reflection of the time behaviour of the EOF maxima, there would be multiple points for each coupling ratio and each curve would therefore be continuous.

\end{document}